# UPGRADE OF PARTICLE FACTORIES

C.Biscari, LNF-INFN, Frascati, Italy

*Abstract*
Recently the luminosity frontier has been raised by the Beauty factories. Higher precision measurements at the energies between the Φ and the Beauty are being projected, and the corresponding increase of luminosity by orders of magnitude is being faced by the accelerator community. The main upgrades and plans in the factories presently in operation or construction around the world are here summarized.

## INTRODUCTION

The lepton colliders are at the frontier of high luminosity. The effort for centre of mass energy increase is now devoted to the hadrons collider construction and facing the era of linear colliders. In the meantime the non-search physics, dedicated to precision measurements, have in the particle factories the field for competitive studies and especially for the accelerator community there is the opportunity of testing new ideas in national laboratory scales.

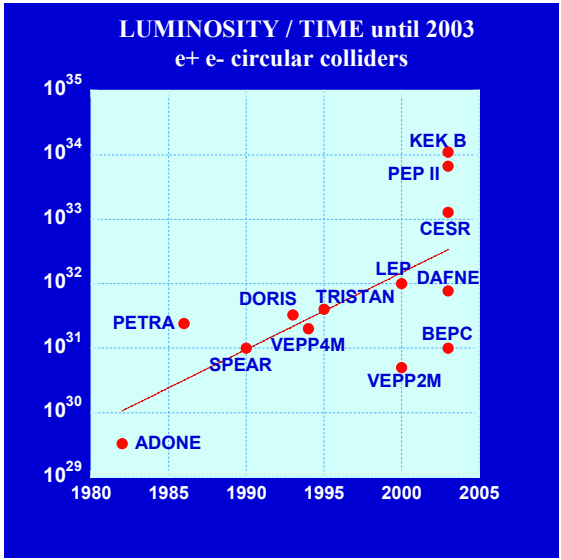

Figure 1 - Lepton collider luminosities versus time.

During the $e^+e^-$ collider history the luminosity has increased in 20 years by 5 orders of magnitude (see Fig.1), and the more spectacular advance has been due to the great success of the B-factories.

For sake of discussion, the diagram of the luminosity, *L*, versus energies, *E*, can be divided in three zones [1] (see Fig.2): in the first zone the energy increase is privileged for new particles production, reaching the maximum with LEP; the corresponding luminosity is continuously improved, both for its dependence on energy and for the advances in technologies and collision physics. Beyond LEP linear colliders will supplant circular ones; VLLC should double LEP energy but at the price of one order of magnitude in ring size. In the second zone there are the present factories, with luminosities one order of magnitude larger than the previous ones at the same intermediate energies. The third zone represents the future, in which upgrades by a factor 10 seem reachable with present technologies, while higher factors are subjected to R&D progress.

The annihilation production cross section in $e^+e^-$ collisions is proportional to the inverse square of the energy, and the necessary integrated luminosity scales accordingly:

$$\int L \propto \frac{1}{\sigma} \propto E^2 \qquad (1)$$

Table I shows the approximate integrated luminosities already collected by all experiments in the energy range between the Φ and the B, and the luminosities requested for competitive experiments in the LHC era. The increase in integrated luminosity should be reached in a reasonable time scale, of few years: very high peak luminosity together with the overall reliability of the collider is the necessary goal for the super-factory projects.

A review of the plans of $e^+e^-$ factories according to their energy is described in the paper.

Table 1 - Collected and requested integrated luminosities

|  | $E_{cm}$ (GeV) | logged $\int L$ | requested $\int L$ |
|---|---|---|---|
| Beauty | 10.6 | ~ 350 fb$^{-1}$ | 10 ab$^{-1}$ |
| τ-charm | 3.9 | < 1 fb$^{-1}$ | >100 fb$^{-1}$ |
| Light Quarks | 1-2 | < 100 pb$^{-1}$ | 500 pb$^{-1}$ |
| Φ | 1 | < 1 fb$^{-1}$ | >100 fb$^{-1}$ |

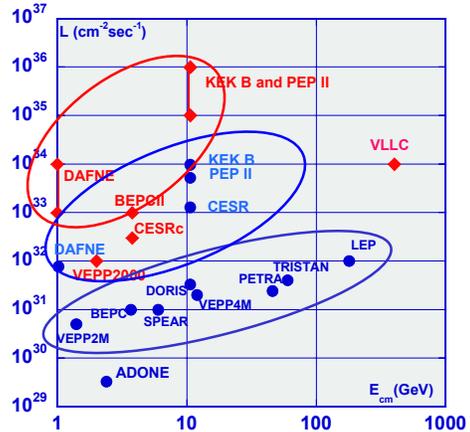

Figure 2: Luminosity versus energy in lepton circular colliders. Past and present results (blue dots), future projects and designs (red diamonds)

# BEAUTY FACTORIES

Two B-factories were conceived in the 90's, with very ambitious design parameters. KEK-B and PEP-II have reached the design aims in a very short time, compared to the scale of accelerator history. PEP-II design luminosity, $3 \cdot 10^{33}$ cm$^{-2}$sec$^{-1}$, was obtained in two years after the start-up (see fig.3), and is now already doubled. The very challenging KEK-B design luminosity, $1 \cdot 10^{34}$ cm$^{-2}$sec$^{-1}$ has been reached few months ago (see fig.4), and is the highest luminosity ever measured. Both colliders are based on similar designs: double rings with asymmetric energies, flat beams and multibunch regime. The beam-beam energy transparency condition, which was one of the design conditions, has been successfully relaxed in course of operation, by finding the best current/beam sizes configuration to optimize luminosity under different regimes. The handling of IR backgrounds, the operation at very high currents, with the successful operation of the bunch-by-bunch feedbacks are a noticeable result. Beam-beam parameters have reached values between 0.06 to 0.09.

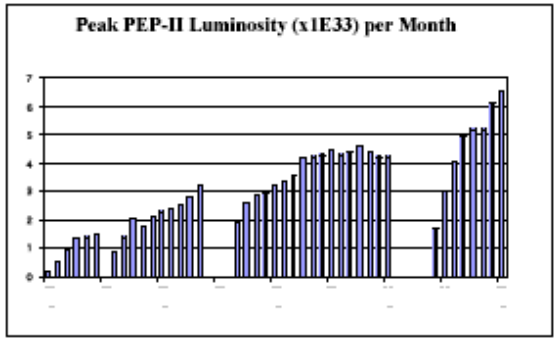

Fig. 3 - PEP-II luminosity

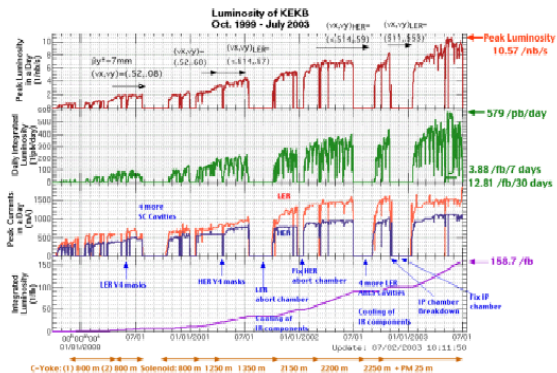

Fig.4 - KEK-B luminosity

CESR has operated in this same range of energy. The collider performances have been continuously improved, to be competitive with the newer collider generations. Now it is operated also at lower energies and can cover all the physics from the J/Ψ to the Beauty.

Both KEKB and PEPII plan upgrades for the next two-three years, optimizing the existing systems to reach luminosities of few $10^{34}$cm$^{-2}$sec$^{-1}$. Further steps on luminosity are being planned in a time scale of 10 years. The main parameters for the future upgrades are summarized in Table2 for both projects [2][3].

Table 2 - B factories from $10^{34}$ to $10^{36}$ cm$^{-2}$sec$^{-1}$

| Collider | KEK-B | | PEP-II | | |
|---|---|---|---|---|---|
| | super | hyper | next | super | hyper |
| $E+$ (GeV) | 3.5 | 3.5 | 3.1 | 3.5 | 3.5 |
| $E-$ (GeV) | 8.0 | 8.0 | 9.0 | 8.0 | 8.0 |
| $C$ (m) | 3016 | 3016 | 2199 | 2199 | 2199 |
| $L\ 10^{34}$cm$^{-2}$s$^{-1}$ | 10 | 40-100 | 2.5 - 4 | 20 | 100 |
| IPs | 1 | 1 | 1 | 1 | 1 |
| $\beta^*$ (m) (h) | 0.30 | 0.15 | 0.5 | 0.3 | 0.15 |
| $\beta^*$ (m) (v) | 0.003 | 0.003 | 0.0065 | 0.0037 | 0.0015 |
| $\varepsilon$ (n rad) (h) | 33 | 33 | 44 | 44 | 44 |
| $\varepsilon$ (n rad) (v) | 2 | 0.33 | 0.44 | 0.44 | 0.44 |
| $\theta$ (mrad) | ±15 | 0 | 0 -±4 | ±10 | ±15 |
| $\xi$ (h) | 0.068 | 0.1 | 0.08 | 0.10 | 0.10 |
| $\xi$ (v) | 0.05 | 0.2 | 0.08 | 0.10 | 0.10 |
| N bunches | 5018 | 5018 | 1700 | 3400 | 7000 |
| $I+$ (A) | 9.4 | 17.2 | 4.5 | 11.0 | 10.3 |
| $I-$ (A) | 4.1 | 7.8 | 2.0 | 4.8 | 2.35 |
| $f_{RF}$ (MHz) | 509 | 509 | 476 | 476 | 952 |

## PEP-II upgrades[2]

The goal of reaching L of the order of $10^{34}$ cm$^{-2}$sec$^{-1}$ by 2005 is based on adding RF stations in order to increase currents and number of bunches and decrease $\beta_y^*$, thanks to the shorter bunch length. An upgrade of the longitudinal feedback system with new electronics and DAΦNE-like kicker will increase the effectiveness of the system (see Fig. 5). Higher injection rate, correlated with added collimators to shield injection background will increase the ratio of average to peak luminosity. Solenoidal windings for ECI together with increased cooling should help in the current increase.

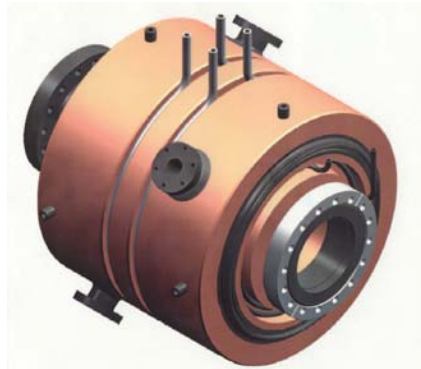

Fig.5 - Longitudinal feedback kicker for PEP-II

The same philosophy is foreseen up to 2008, but pushing parameters to more limiting values: $\beta_y^*$ will be decreased by 50% by moving quadrupoles closer to IP, a small crossing angle in the new IR will be introduced, higher currents will be based on the feedback system upgrade to go to 2-bucket spacing. All these actions should push luminosities up to 2-4 $10^{34}$ cm$^{-2}$sec$^{-1}$. $\xi_{x,y}$ of 0.08 are considered achievable.

Higher luminosity considerations will be of course related to the achievements obtained at that point; with the today know-how, the idea is to increase the collision frequency by filling all the buckets, without increasing bunch currents, together with a larger crossing angle and smaller betas at the IP. The total current will be doubled and to save wall power the energy asymmetry will be diminished. Tune shifts will be 0.1 and $L$ of the order of $10^{35}$cm$^{-2}$sec$^{-1}$.

The main upgrade for a further increase is the change of the RF frequency by a factor of two to double the number of bunches. With $f_{rf}$ equal to 950 MHz, the bunch separation goes to 1 nsec, and a further increase of the crossing angle is foreseen. Figure 6 shows the IR layout for this configuration. With the same b-b tune shift and slightly lower currents per bunch the total $L$ could reach values of $10^{36}$ cm$^{-2}$sec$^{-1}$. R&D on the RF cavity and related systems is already in progress.

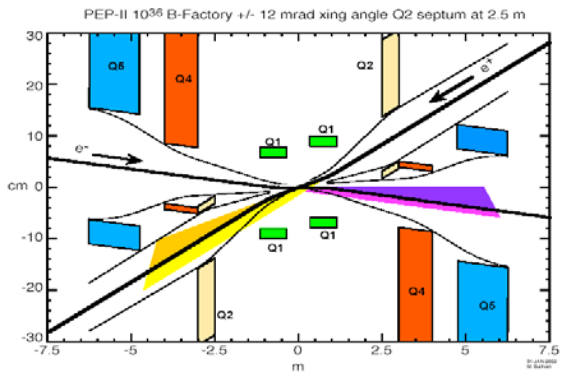

Fig.6 - PEP-II IR for L = 1 $10^{36}$ cm$^{-2}$sec$^{-1}$

The beam lifetimes will be very low and continuous injection will be needed and it will be used to push the beam-beam parameter to higher values than those which can be tolerated when long lifetimes are required. Bunch-by-bunch feedbacks will need to operate at the 1 nsec scale, down from the present 4 nsec time. Much shorter bunches will be needed, of the order of 2 mm. Higher-power vacuum chambers and HOM tolerant chambers will be needed. The use of expansion bellows will need to be minimized or a high-power design developed. Very low vertical beta functions at the interaction of about 1.5 to 2.5 mm will be needed, together with special chromaticity corrections. Every technique to reduce the wall plug power will be used: for example, increase the vacuum chamber bores to reduce resistive wall effects, and increasing the RF cavity bores to reduce HOM losses.

## KEK-B upgrades[3]

The Japanese B-factory exceeded the $10^{34}$ cm$^{-2}$sec$^{-1}$, the maximum luminosity ever reached, in May 2003. Not only the peak luminosity, but also the continuous reliability of the whole collider, has allowed the BELLE detector to collect a huge amount of data in a very short time. At the end of this year the collider runs normally at peak luminosity higher than $10^{34}$ cm$^{-2}$sec$^{-1}$ and integrated luminosity per day exceeding 600 pbarn$^{-1}$.

The upgrades for a factor 10 on the luminosity are based on an increase of the colliding current by multiplying the bunch number by a factor of 4, and increasing also each bunch current, lowering $\beta_y^*$ by a factor of 2 and increasing the crossing angle. By increasing the horizontal emittance the b-b tune shifts are maintained equal to the present ones. The main challenge is the high current effects; the RF system will be upgraded and SC RF cavities will be added (see fig. 7)

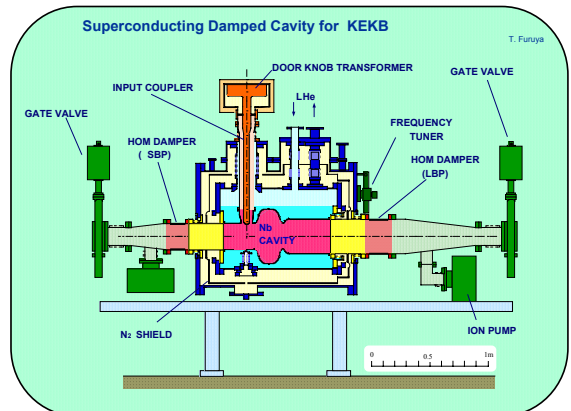

Fig. 7 - Superconducting RF cavity for Super KEKB

Since the electron cloud effect is one of the strongest limit to the increase of the current in the low energy ring, where more current must be stored to keep the transparency condition, a beam-energy switch is envisaged, to store e$^+$ in the high energy ring. An upgrade of the injector is thus being studied to accelerate positrons up to 8 GeV. Intensive R&D on vacuum chamber design, with antechambers and special RF shields is in progress, and prototypes are being constructed to be tested soon in the collider.

Very high beam-beam parameter, >0.1, is needed to get the high luminosity, otherwise very high operating current or very small beta function at the collision point are required. Beam-beam simulations are being carried out; they indicate that at very high values of tune shift the effects of crossing angle are dramatic. Based on these results the philosophy for reaching $10^{36}$ cm$^{-2}$sec$^{-1}$ is based on head-on collisions obtained with crab cavities. The first test crab-cavity will be installed in one year in the ring for experiments. Lowering the coupling should help in obtaining the very high $\xi_y$.

# TAU-CHARM FACTORIES

## BEPCII, CESRc

CESR[4] has been upgraded to operate at lower energies. The lengthening of the damping time is fought by new wigglers (see fig.8) which increase radiation damping. The first six wigglers have been already installed and commissioned. Other six wigglers will be installed in one year and CESRc will run until 2008 at three energies between 3.1 and 4.1 GeV.

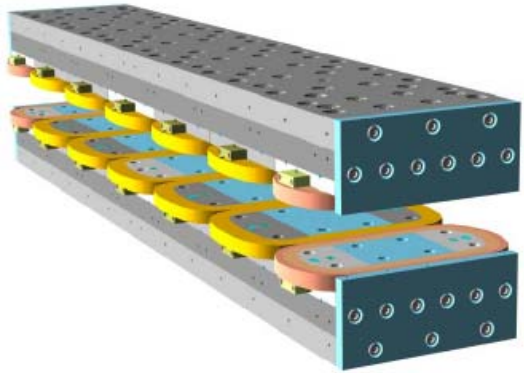

Fig. 8 - CESR-c wiggler

The Chinese collider BEPC[5] is being upgraded to become the first completely dedicated tau-charm factory, still maintaining the synchrotron radiation production. Its design is based on the double ring scheme (see fig.9), with energies ranging between 1.5 and 2.5GeV per beam, optimized at 1.89 GeV. An inner ring will be installed inside the old one, so that each beam will travel in half outer ring and half inner one. Superconducting cavities fitting the bunch length requirements will be installed. The production began in 2002 and commissioning is foreseen for 2006.

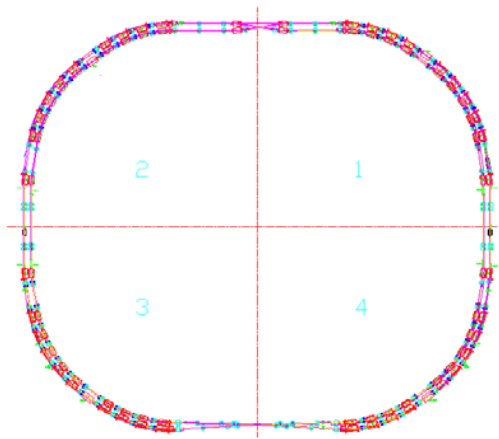

Fig.9 - BEPC layout

Table 3 shows the design values of the Chinese τ-charm factory together with the CESR-c parameters at the same energy. The status of both projects is well illustrated in these conference proceedings [6,7].

Table 3 – τ - charm factories

| Collider | CESRc | BEPC II |
|---|---|---|
| status | operating | in construction |
| $E$ (GeV) | 1.88 | 1.89 |
| $C$ (m) | 768 | 237.5 |
| $L$ ($10^{32}$ cm$^{-2}$s$^{-1}$) | 3 | 10 |
| IPs | 1 | 1 |
| $\beta^*$ (m) (h / v) | 0.7/ 0.011 | 1 / .015 |
| $\varepsilon$ (μ rad) (h / v) | 0.22 | 0.17 / 0.002 |
| $\theta$ (mrad) | ± 2.8 | ± 11 |
| $\phi$ (rad) | 0.07 | 0.4 |
| $\sigma_z$ (cm) | 1.0 | 1.5 |
| $N_b$ ($10^{10}$) | 6.4 | 4.8 |
| $\xi$ (h / v) | 0.03 / 0.03 | 0.04 / 0.04 |
| N bunches | 45 | 93 |
| I (A) | .18 | 0.91 |
| $f_{RF}$ (MHz) | 500.0 | 499.8 |
| V (MV) | 10 | 1.5 |

# LIGHT QUARKS FACTORIES

Physics at energies between the φ and the τ−charm has been investigated during last years in VEPP-2M (shut-down in 2000) and BEPC colliders. The interest for this energy range had produced the proposal for PEP-N [8]. Now a collider, innovative in its design and concepts, is in construction: VEPP2000. At this energy range the interesting physics needs moderate integrated luminosities, as shown in Table 1.

## VEPP2000

A 2 GeV collider (from there the 2000 in the name), whose design is based on the concept of round colliding beams, is being constructed in Novosibirsk[9], after the shutdown of VEPP-2M three years ago. This is a very important step in the beam-beam interaction understanding. The expected b-b tune shift is half the corresponding flat-beam one with the same particle density, thus predicting a single bunch luminosity of $10^{32}$cm$^{-2}$sec$^{-1}$.

The collider can be operated also with flat beams and at energies ranging from 500 Mev to 1 GeV per beam. Its compact design is based on very high field normal conducting dipoles (2.4T) and houses two experiments in the two symmetric Interaction Regions. Focusing in the two interaction regions is performed by SC solenoids, which also rotate by π/2 the planes of betatron oscillations, thus originating emittance in both transverse modes. Dynamic aperture is challenging due to the high chromaticity and beam sizes on both planes.

Dipoles are being installed, solenoids are in the construction phase and first beam is foreseen at the end of next year.

The full description of the project is the subject of a contribution to the workshop [9]. Figure 10 shows an artistic view of the ring.

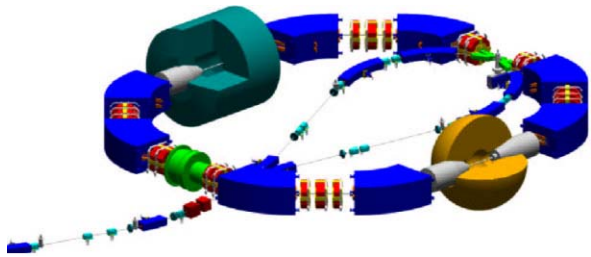

Fig. 10 - View of the VEPP-2000 collider

Table 4 - Light quarks factories

| Collider | VEPP2000 | DAFNE 2 |
|---|---|---|
| status | in construction | design study |
| E (GeV) | 1. | 1. |
| C (m) | 24 | 97 |
| L ($10^{32}$ cm$^{-2}$ s$^{-1}$) | 1 | 1 |
| IPs | 2 | 1 |
| $\beta^*$ (m) (h / v) | 0.1 / 0.1 | 1.5 / 0.025 |
| $\varepsilon$ ($\mu$ rad) (h / v) | 0.136 / 0.136 | 0.5 / 0.0025 |
| $\theta$ (mrad) | 0 | ±15 |
| $\phi$ (rad)(Piw) | 0 | 0.26 |
| $\sigma_z$ (cm) | 3 | 1.1 |
| $N_b$ ($10^{10}$) | 10 | 3 |
| $\xi$ (h / v) | 0.1 / 0.1 | 0.014 / 0.024 |
| N bunches | 1 | 30 |
| I (A) | 0.20 | 0.45 |
| $f_{RF}$ (MHz) | 172 | 368.3 |
| V (MV) | 0.12 | 0.25 |

### DAFNE2

DAΦNE has been constructed to operate at the Φ resonance. Some of the systems are dimensioned to operate also at higher energies. It is presently under discussion which will be the future of the collider. One of the possibilities is to increase the energy by a factor of two [10]. The project is named DAFNE2, where "F" stands for Frascati and "2" for $E_{cm}$. No crucial issues from the accelerator physics point of view are envisaged. The main hardware modifications concern the dipoles and the Interaction Region, while RF and vacuum systems are already dimensioned for the high energy, with a lower beam current. The main parameters are given in table 4 together with those of VEPP2000 for comparison. To be noticed that the same luminosity claimed by both projects has an essential difference: in VEPP2000 it is obtained in single bunch mode, while in DAFNE2 is obtained in multibunch configuration.

## Φ-FACTORIES

DAΦNE is the only Φ-Factory presently in operation. VEPP2000 can be operated also at the Φ resonance, with of course luminosity lower than the optimum one.

### DAΦNE

In two or three years all the DAΦNE current physics programs are expected to be completed, with an overall delivered integrated luminosity in excess of 3fb$^{-1}$ and luminosities higher than $10^{32}$cm$^{-2}$sec$^{-1}$. [11]

The interest for values of the luminosity larger by a factor 10 than the design ones has led to the study of possible new designs of the factory.

During this workshop the design of a Φ-factory based on the strong RF focusing principle is discussed in several contributions (see for example [12], [13]). At the state of art, this idea seems to be the more appealing in terms of luminosity increase for the low energy factory.

## CONCLUSIONS

In the next future the present lepton factories will be optimized by stretching the design parameters, without introducing essentially any new concept in the luminosity production, expecting in all cases an increase in peak and integrated luminosity by about a factor 10.

New ideas to push the luminosity values by another order of magnitude are being investigated. Round beam collisions will soon be tested at VEPP2000, answering to the question whether b-b tune shift limits can be raised. The strong rf focusing principle, introduced as a possibility to obtain very short bunches at the IP, deserves a proof of principle for validation.

## REFERENCES

[1] C.Biscari, "Future Plans for e+e- Factories" – Proceedings of PAC 2003, Portland, May 2003
[2] J.Seeman et al, "Design Studies for a $10^{36}$ SuperB-Factory" – 31$^{th}$ ICFA Beam Dynamics Newsletter, p.72 – August 2003
[3] K. Ohmi, M. Tawada, "Challenge Toward Very High Luminosity At Super KEKB", 31$^{th}$ ICFA Beam Dynamics Newsletter, p.95 – August 2003
[4] D. L. Rubin, M.J. Forster, "CESR-c Lattice Design and Optimization", 31$^{th}$ ICFA Beam Dynamics Newsletter, p.43 – August 2003
[5] P.D. Gu et al., "Physics Design of BEPCII", 31$^{th}$ ICFA Beam Dynamics Newsletter, p.32 – August 2003
[6] D.Rubin et al, "CESR-c", these proceedings
[7] C.Zhang et al, "BEPCII: Status and Progress", these proceedings.
[8] Y.Cai et al. "PEP-N: a 0.8GeV x 3.1 GeV Collider at SLAC", Proc. of PAC2001, Chicago, p.3564
[9] Y.Shatunov et al. " Status of the VEPP-2000 Collider Project", these proceedings.
[10] G. Benedetti et al., " Feasibility study of a 2 GeV lepton collider at DAΦNE", Proceedings of PAC 2003, Portland, May 2003
[11] M. Zobov, "Beam Dynamics Issues in DAΦNE e+e- Φ-Factory", 31$^{th}$ ICFA Beam Dynamics Newsletter, p.14 – August 2003
[12] A.Gallo, P.Raimondi, M.Zobov, "Strong RF Focusing for Luminosity Increase", these proceedings.
[13] C.Biscari, "Lattice for Longitudinal Low-Beta", these proceedings